\documentclass[12pt]{article}

\usepackage{scicite}
\usepackage{graphicx}
\usepackage{times}
\usepackage{amsmath}
\usepackage{xcolor}
\makeatletter\def\author#1{\gdef\@author{\hskip-\dimexpr(\tabcolsep)\hskip1pt\parbox{\dimexpr\textwidth-1pt}{\centering#1}}}\makeatother

\newenvironment{sciabstract}{%
	\begin{quote} \bf}
	{\end{quote}}

\title{Human body heat shapes the pattern of indoor disease transmission}

\author
{Chao-Ben Zhao$^{1}$, Jian-Zhao Wu$^{1}$, Bo-Fu Wang$^{1,2}$, Tienchong Chang$^{1,2}$,
Quan Zhou$^{1,2}$ and Kai Leong Chong$^{1,2\ast}$\\
~\\
\normalsize{$^{1}$Shanghai Key Laboratory of Mechanics in Energy Engineering, Shanghai Institute of Applied Mathematics and Mechanics, School of Mechanics and Engineering Science, Shanghai University,}\\
\normalsize{Shanghai 200072, PR China}\\
\normalsize{$^{2}$Shanghai Institute of Aircraft Mechanics and Control,}\\
\normalsize{Zhangwu Road, Shanghai 200092, China}\\	
~\\
\normalsize{$^\ast$To whom correspondence should be addressed; E-mail: klchong@shu.edu.cn}\\
~\\
\normalsize{\textbf{A one-sentence summary of the paper:} \textit{ Our study uncovered two thermal plume regimes---individual and collective---and a critical distancing threshold that triggers a transition between them, revealing novel disease spread patterns through the resulting morphological phase transition of airflow.}}
}

\date{}

\begin{document} 
	
	
	\baselineskip24pt
	
	
\maketitle

\begin{sciabstract}

Exhaled droplet and aerosol-mediated transmission of respiratory diseases, including SARS-CoV-2, is exacerbated in poorly ventilated environments where body heat-driven airflow prevails. Employing large-scale simulations, we reveal how the human body heat can potentially spread pathogenic species between occupants in a room. Morphological phase transition in airflow takes place as the distance between human heat sources is varied which shapes novel patterns of disease transmission: For sufficiently large distance, individual buoyant plume creates a natural barrier, forming a ``thermal armour'' that blocks suspension spread between occupants. However, for small distances, collective effect emerges and thermal plumes condense into super-structure, facilitating long-distance suspension transport via crossing between convection rolls.  Our quantitative analysis demonstrates that infection risk increases significantly at critical distances due to collective behavior and phase transition. This highlights the importance of maintaining reasonable social distancing indoors to minimize viral particle transmission and offers new insights into the critical behavior of pathogen spread.

\end{sciabstract}
    
\section*
  {Introduction}
Outbreaks of respiratory diseases, such as influenza, severe acute respiratory syndrome (SARS), middle East respiratory syndrome, and now the SARS-CoV-2, have taken a heavy toll on human populations worldwide. Until vaccines are widely available, it is commonly held that the useful precautions are case isolation, wearing mask, social distancing\cite{Bourouiba_jama_2020,mittal_jfm_2020,Ku_SciAdv_2021,Yafang_sci_2021,Pauser_scir_2021,Leung_nrm_2021}, which relate closely on the flow physics of respiratory suspensions\cite{Lidia_ei_2020,Gholamhossein_pnas_2021,Adam_na_2022}. Nowadays, flattening the curve is an important public health strategy to cope with the pandemic. To efficiently slow down the spread of disease, a relevant question to be answered is how people maintain a safe distance with an infected person as they interact together in an enclosed room, which is especially important as we would gradually return to the normal social interaction in post-pandemic phase. In contrast to outdoor events where dilution of contaminant can be brought about by  wind, indoor social interactions between human bodies, however, could lead to a significant risk for the spread of SARS-CoV-2, especially when the ventilation is inadequate\cite{Miller_idm_2018, Kimberly_sci_2020,Morawska_cid_2020,Guo_Eeid_2020,Wenzhao_be_2020,Miller_ia_2021,Seo_iji_2021,Mathai_sciacv_2021}. In indoor situations, saliva droplets or aerosols are released into the environment by asymptomatic or presymptomatic infected individuals\cite{wells_aje_1934,Wells_jama_1936,xie_ia_2007,Chan_Lancet_2020,Hu_scicn_2020,Rothe_njm_2020,Abkarian_prf_2020,nordsiek_po_2021}, then suspended and mixed by the airflow for hours or even days \cite{Abkarian_pnans_2020,Yang_prf_2020,Chong_prl_2021,Bourouiba_jama_2020,Justin_Science_2021,Hayden_sica_2021}. This kind of indirect transmission involves infections in large spatial and temporal scales, and possibly leads to the superspreading events\cite{Althouse_pb_2020,Young_cell_2020,Li_bae_2021,Qian_ia_2021}.

Surprisingly, little is known on long-time dispersion and transmission pattern of suspended viral particles in indoor environment. 
Such physical process of spreading is subtle as it involves interactions between respiratory droplets, turbulent eddies, vapor and temperature fields\cite{Bourouiba_jfm_2014,Bourouiba_jama_2020,Balachandar_ijmf_2020,Bourouiba_arfm_2021,Chong_prl_2021,Ng_prf_2021,Jietuo_pnas_2021}, and also relates to several environmental factors such as ventilation\cite{Bhagat_jfm_2020,Wen_bae_2020,Lidia_sci_2021,yang_jfm_2022}, ambient temperature\cite{lin_jfm_2006,Anice_jof_2008} and relative humidity\cite{yang_PLoS_2011,Marr_jos_2019,lin_Lin_2020,Walker_ASC_2021,Aganovic_sciR_2022}. In practice, ventilation is considered as an active control strategy to transport and exhaust those viral aerosols or droplets. 
For the sake of carbon dioxide removal and energy saving, the ventilating flow needs not to be too strong except in some dedicated cases. For example, the average flow  speed is $4-8$mm/s for the ventilation rate of  $N=5-10$ACH\cite{Bhagat_jfm_2020}. However, in such situations, the thermal plumes emitting from a human body  becomes significant to drive the indoor flow and thus spread the droplet nuclei in enclosed spaces. Considering the body-and-ambient temperture difference of $8^o$C, the speed of thermal plume generated from a person is approximately 800mm/s using the characteristic buoyancy veloctiy scale, giving the flow even much stronger than the normal indoor ventilating flow. Hence, the body heat source and its distribution can crucially shape the pattern of viral transmission in indoor environment.

To address the question of how the body heat drives the flow pattern and affects the disease transmission, we carry out a series of large-scale simulations of indoor flow to examine the aerosol transport with different spacings between the heat source, mimicking different social distancing. 
We discover a transition in flow structure from individual to collective and condensed thermal plumes as heat-carrying body density increased, shaping different disease transmission patterns. 
In individual regime, the so-called thermal armour is formed which traps the aerosols in localized circulation. On the contrary, in collective regime, merging of plumes enables the long-distance aerosol transport. We evaluated the risk of cross-infection at different spacings in poorly ventilated rooms.

\begin{figure*}[htbp]
   \centering
   \setlength{\unitlength}{\textwidth}
   \includegraphics[width=1.0\unitlength]{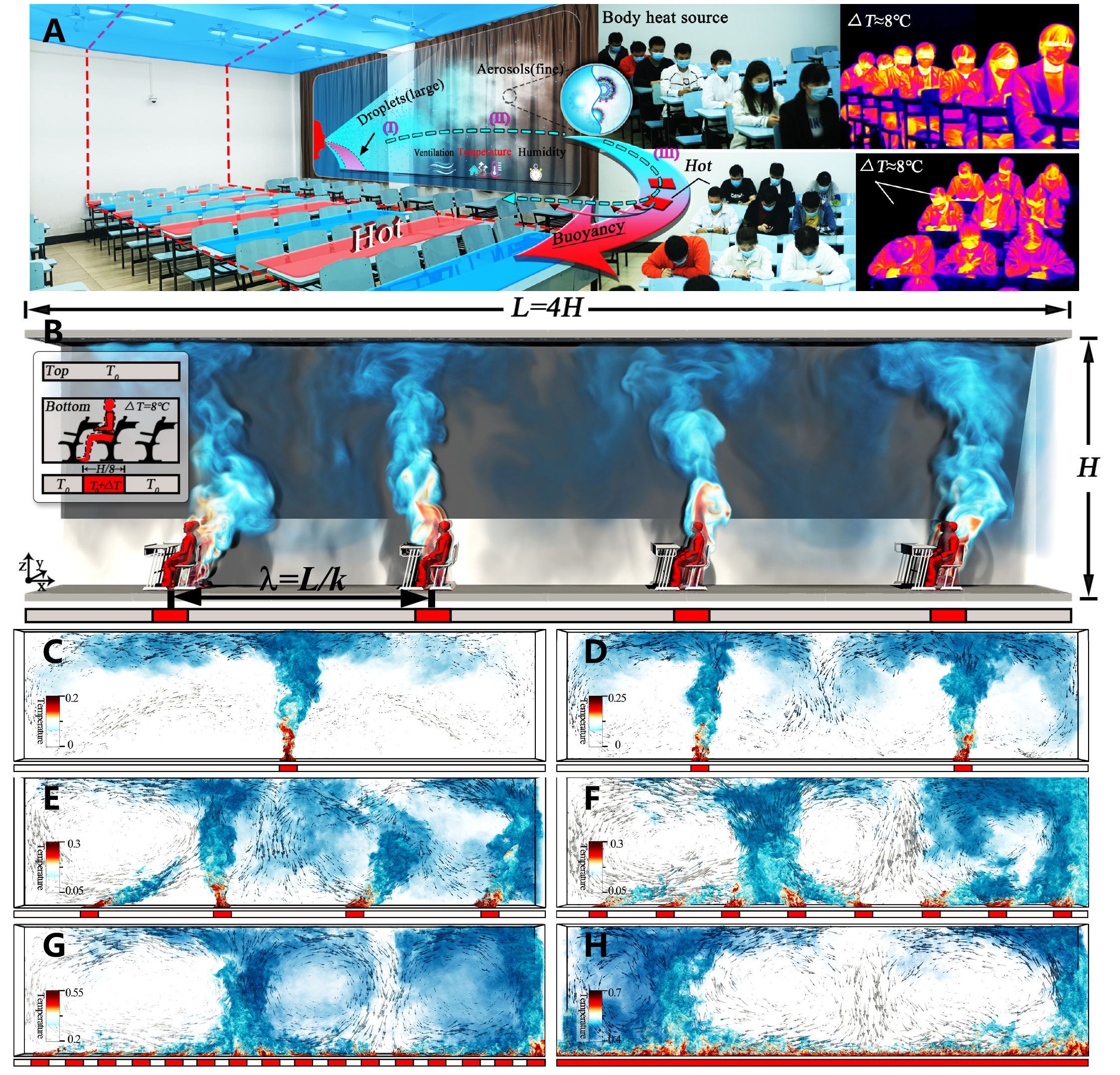}
	\caption{\textbf{Airflow patterns driven by human body plumes.} \textbf{A}, Major modes of transmission of respiratory viruses during short-range and long-range transmission.  During respiratory infections, individuals expel virus-laden droplets and aerosols while breathing. Large droplets settle quickly (I), but aerosols persist in the air, increasing transmission risk in poorly ventilated areas (II).  Infrared cameras detect thermal emissions in enclosed spaces like classrooms, revealing elevated body surface temperatures with a distinct orange-red hue (III). \textbf{B}, Illustration of the simulation set-up with the heating source of human body. \textbf{C}-\textbf{H}, The snapshot of the temperature (color) and velocity (arrows) field in the steady state for the different distancings  of people in the room: $\lambda=L$ in \textbf{C}, $\lambda=L/2$ in \textbf{D}, $\lambda=L/4$ in \textbf{E}, $\lambda=L/8$ in \textbf{F}, $\lambda=L/16$ in \textbf{G}, $\lambda=L/32$ in \textbf{H}.}
	\label{fig:sketch}
\end{figure*}

\section*
  {Results}
\subsection*{Overall airflow patterns}

We consider a room of dimensions $L:W:H = 12m \times 1m \times 3m$, as sketched in Fig.~\ref{fig:sketch}B, with a narrowed room being considered. An occupant sitting in the room produces thermal plumes due to the temperature contrast between the ambient and human body. For simplicity, the heat source of the seated occupant is evenly distributed, and the ambient temperature of the room is considered to be homogeneous initially and set to the value of normal indoor temperature, i.e. $T_{amb}$ = 26$^{\circ}$C. We simulate the buoyancy-driven effect of the body heating by setting the temperature boundary conditions. The heating temperature of the body is set to be $T_{ocpt}$=34$^{\circ}$C and the width of the heat source to be 0.375m. The fluid is assumed to be incompressible $\nabla \cdot \boldsymbol{u} = 0$. The gas Prandtl number is $Pr\equiv \nu /\kappa_T= 0.7$, where $\nu$ is the kinetic viscosity of air and $\kappa_T$ is the thermal conductivity. The thermal Rayleigh number defined using the occupant height $h_{ocpt} = 1.2 m$ is $Ra_{ocpt}\equiv \alpha g \Delta h^3_{ocpt}/(\nu \kappa_T)\approx 10^9 $, where $\Delta \equiv T_{ocpt} -T_{amb}=8^{\circ}C$ \cite{yang_jfm_2022}, and $g$ is the gravitational acceleration. We characterize the spacing of human heat sources by $\lambda$, which equals to the length $L$ divided by the number of human body $k$ ($\lambda = L/k$), which is also an important control parameter of the system. Further details of control parameters are given in the Supplementary Materials.

The flow structure driven by regularly arranged occupants in a poorly ventilated room is first examined. Figure~\ref{fig:sketch}C-H shows the typical instantaneous flow structure at different distances $\lambda$ of body heating sources, visualized by the volume rendering of temperature field. 
It is clearly seen that the fluctuating thermal plume is emitted from the heating source and rises into the ambient air. For large distance $\lambda$, \textit{i.e.} with small number of occupants as shown in Fig.~\ref{fig:sketch}C-E, those ejected plumes self-organize into the circulating flow (see supplementary Movie. S1). The interior spaces of room are then separated by the airflow pattern formed by several individual circulations. 
For small distance $\lambda$, \textit{i.e.} with large number of occupants as shown in Fig.~\ref{fig:sketch}F-H, the collective buoyancy effect is found, where the large-scale flow induces strong shear effect to favour the merging of thermal plumes (see supplementary Movie. S2). The condensation of plumes connects a number of heating sources, resulting in the super-structure in the enclosed room.

\begin{figure}[!hbt]
	\centerline{\includegraphics[width=1.0\linewidth]{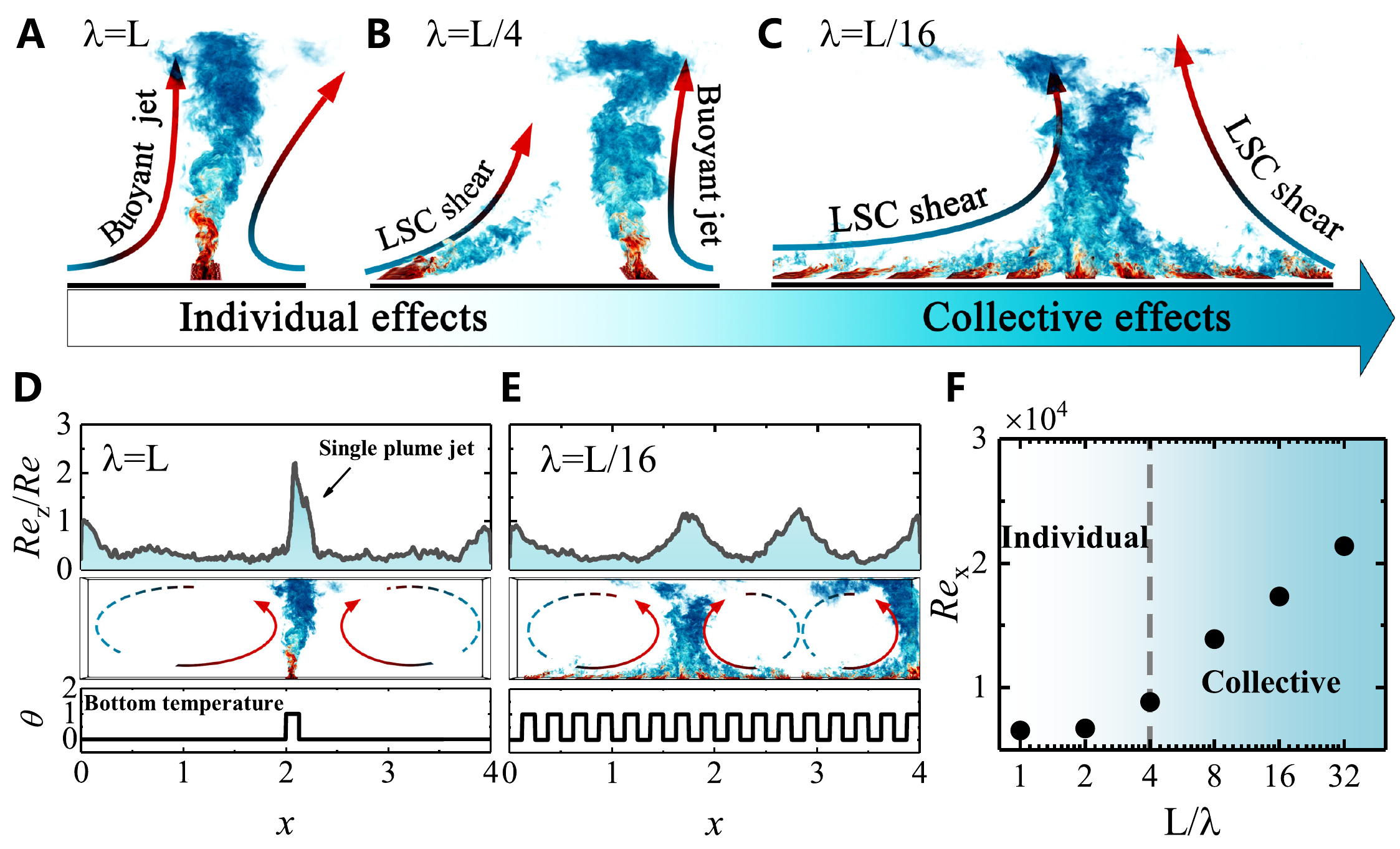}}
	\caption{\textbf{Individual and collective effects of body plume motions}. \textbf{A}-\textbf{C}, Behavior of body plumes for different distances between occupants: $\lambda=L$ in \textbf{A}, $\lambda=L/4$ in \textbf{B}, $\lambda=L/16$ in \textbf{C}. \textbf{D}, \textbf{E}, The Reynolds number ratio $Re_z/Re$ as a function of the horizontal position $x$ obtained from the slice of the velocity field at mid-width. Here $Re_z=\sqrt{\langle w^2 \rangle_{z}}H / \nu $ is the averaged vertical Reynolds number, $Re=\sqrt{\langle u^2+v^2+w^2 \rangle_{z}}H / \nu$ where $\langle \cdot \rangle_{z}$ denoting the averaging over vertical direction. As shown in \textbf{D} at $\lambda = L$, the plume emission at the peak position leads to local flow enhancement, showing significant individual effects. However, at $\lambda = L/16$ in \textbf{E}, the location of the plume-rising area does not entirely depend on the individual heat source, the behavior of the plume shifts to a collective effect. \textbf{F}, The horizontal Reynolds number $Re_x$ as a function of the ratio between the length andx distancing $L/\lambda$. Here $Re_x=\sqrt{\langle u^2 \rangle_{V,t}}H \nu $ is the spatially and temporally averaged horizontal Reynolds number.}
	\label{fig2}	
\end{figure} 

The overall airflow pattern undergoes a morphological transition from the individual to the collective behavior of thermal plumes as the distance between occupants decreases as shown in Fig.~\ref{fig2}A-C. At $\lambda=L$, the motion of plume exhibits strong individual effect, \textit{i.e.}, thermal plume from heating source directly moves upward without interaction with the adjacent plumes. With decreasing the distance $\lambda$ (e.g. $\lambda=L/4$ and $\lambda=L/16$), body plumes start to merge with each other as favoured by the shearing of the large-scale mean flow, and eventually the strong collective behaviour appears with all thermal plumes contribute energy to a single large-scale flow.

To quantitatively distinguish the individual and collective effects, we plot in Fig.~\ref{fig2}D,E the profiles of $Re_z/Re$ along the horizontal direction, which indicates the strength of buoyancy-driven jet from the heat source. Here, $Re_z$ is the Reynolds number defined by the vertical velocity magnitude, $Re$ is the overall Reynolds number defined by flow speed considering all velocity components. At $\lambda=H$ in the individual regime (Fig.~\ref{fig2}D), the location of the peak coincides with the location of the heat source, and the value of the local $Re_z$ is approximately twice of the overall Reynolds number $Re$. This indicates that the airflow for small number of occupants (in individual regime) is mainly contributed by the buoyancy-driven jet with the dominance of the vertical momentum. At $\lambda=L/16$ under the collective regime (Fig.~\ref{fig2}E), the location of the plume-rising area does not have one-to-one correspondence with the individual heat sources due to merging of plumes. In this case, the horizontal momentum becomes dominant with the emerged super-structure as shown in Fig.~\ref{fig2}F. The consequence of the enhanced horizontal momentum is the potential transport of viral aerosols over a long lateral distance.

\begin{figure}[!hbt]
	\centering\includegraphics[width=1.0\linewidth]{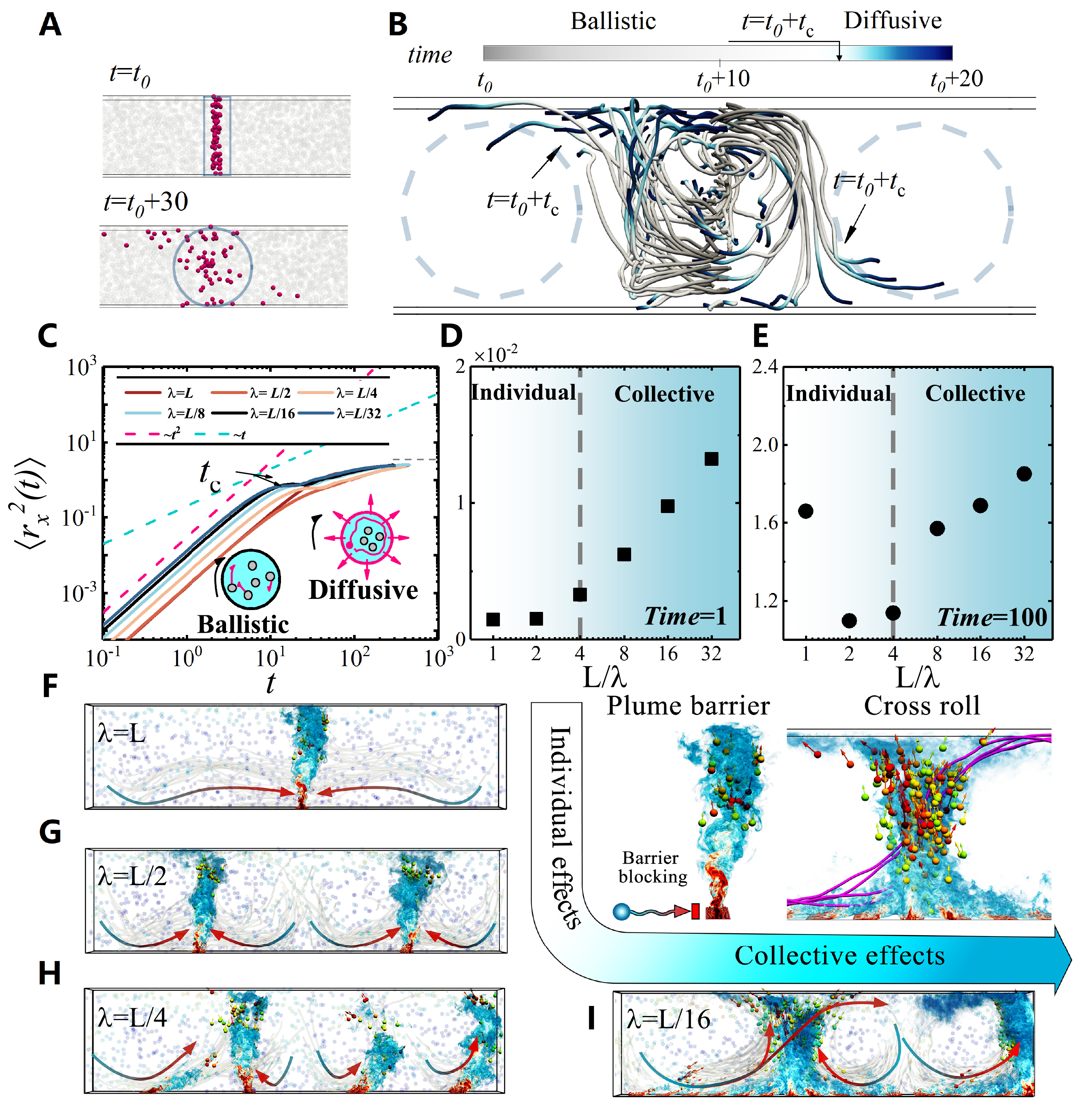}
	\caption{\textbf{Transport characteristics of aerosol transport.} \textbf{A}, Ballistic and diffusive motion of the particles. Typical spatial distribution of aerosols in the ballistic and the diffusive regime. \textbf{B}, Examples of particle trajectories for both ballistic and diffusive motion. \textbf{C}, The mean-squared displacement (MSD) $\langle {r_x}^2(t) \rangle$ of particles as a function of time. The straight dotted lines indicate the theoretical scaling laws corresponding to ballistic and diffusive regimes. \textbf{D},\textbf{E}, The measured MSD as a function of $L/\lambda$ at $t=1$ in \textbf{D} and $t=100$ in \textbf{E}. \textbf{F}-\textbf{I}, Instantaneous spatial distributions of particles overlaid with the thermal plume.}
	\label{fig:MSD}	
\end{figure}

\subsection*{Viral transmission analysis}

It is highly desirable to understand how the airflow transition affect the aerosol transportation, which is important to disease transmission. As most of exhaled aerosols are smaller than $5 \mu m$ \cite{Wangsic_sci_2021,Sima_ast_2020}, and a large fraction is $<1 \mu m$ for most respiratory activities, including those produced during breathing, talking, and coughing, we employ the Lagrangian tracking approach to simulate the trajectories of the floating aerosols. The effect of heat sources on viral particle transmission is examined through statistical analysis on the trajectory of aerosols. 
As the size of aerosols is small enough, they can be regarded as faithful tracers of the fluid flow\cite{Warhaft_arfm_2000,Microbubbles_prl_2016,Mathai_sciacv_2021}.

The positions of thousands of Lagrangian tracer particles are initialized randomly over the domain at a time when the turbulent flow has attained a statistically stationary state. With the obtained trajectories, the statistical behavior of particles can be characterized by their mean-squared displacement (MSD). Figure \ref{fig:MSD}C shows the MSD versus time $t$ for various distance $\lambda$. The trend of MSDs for different $\lambda$ are seen to exhibit similar behavior, namely, at short time, the particle motion is ballistic ($\langle {\vec{r}}^2(t) \rangle \sim t^2$) since the aerosols are advected by the circulation (Fig.~\ref{fig:MSD}A,B); after a certain time $t_c$, a short range of diffusive motion ($\langle {\vec{r}}^2(t) \rangle \sim t^1$) is found and eventually the MSD becomes independent of $t$ due to the finite-size effect. 

To quantify how rapidly the aerosols spread inside the room, we plot in Fig.~\ref{fig:MSD}D, E the magnitude of MSD as a function of $1/\lambda$ at two time instants $t=1$ and $t=100$ (\textit{i.e.} about $3$ secs and $6$ mins in real time). 
At $t=1$ where the motion of aerosols is still within the ballistic stage, the trend of displacement in horizontal direction $\langle r_x^2(t) \rangle$ is similar to that of $Re_x$ (see Fig.~\ref{fig2}F) because the amount of displacement is set by the strength of the flow. At $t=100$ in diffusive stage (see Fig.~\ref{fig:MSD}D), the transition from individual to collective regime can be clearly distinguished: the horizontal displacement decreases with increasing $\lambda$ in the individual regime, and then increases with $\lambda$ in the collective regime. 

This above trend can be understood from the following visualizations. Figure~\ref{fig:MSD}F-I presents the snapshots of the spatial distribution of particles overlapped with the contour of temperature field. In the individual regime, as illustrated in Fig.~\ref{fig:MSD}F-H, the thermal plumes from heating source behave as ``thermal armour''. Consequently, the movement of particles is restricted within the circulation in between ``thermal armour'' and thus particle horizontal displacement decreases with heat source density. However, the effect of ``thermal armour'' has been broken down once the collective regime enters with the merging of plumes taking place. With  plume merging, the airflow in the room becomes dominated by the super-structure, enabling a much further horizontal transport. Since the flow intensity increases with larger number of heat source, there is also opportunity for particles to travel longer distances by crossing several super-structure (Fig.~\ref{fig:MSD}I).

\begin{figure*}
	\centering\includegraphics[width=1.0\linewidth]{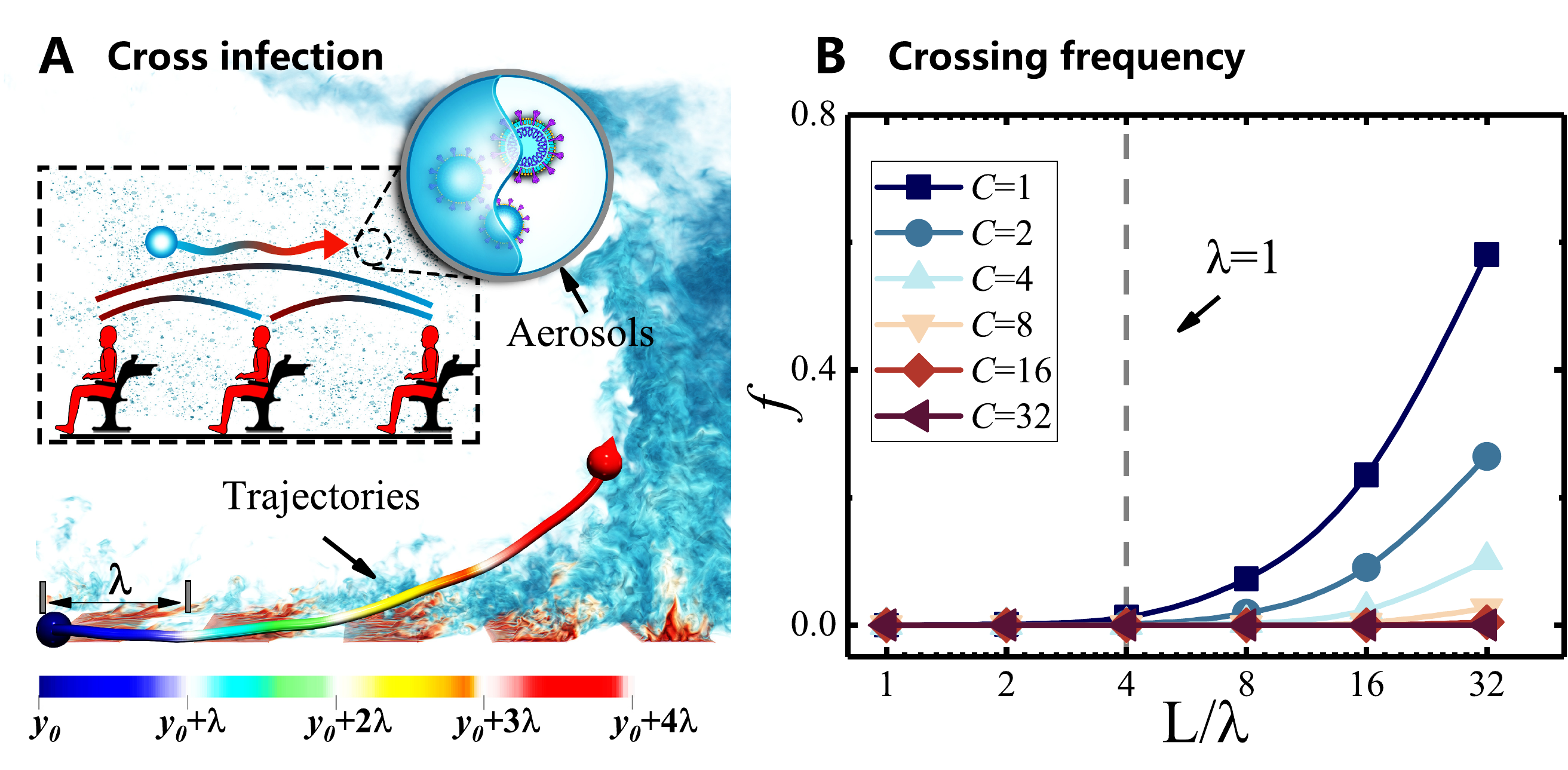}
	\caption{\textbf{Infection risk assessment.} \textbf{A}, Cross infection path: the flows established through the large-scale circulation provide a path for air transmission between the two occupants, and hence a possible infection route. \textbf{B}, The frequency of viral particles crossing over multiple heat sources as a function of $L/\lambda$.  Here, the frequency is calculated by $f=\frac{1}{T}\frac{1}{N}\sum\limits_{i}A_{i}(C)$, where $A_{i}(C)$ is the counts of viral particles released from a certain occupant crossing over multiple neighbor occupants of number $C$, $N$ the total number of particles during a large duration $T$. Once the horizontal displacement of the particles exceeds the distance between the heat sources, we count once on $A_{i}(C)$, \textit{i.e.}, we execute the operation $A_i(C) = A_i(C) +1$ once $\vert X_{i,t+\tau}- X_{i,t} \vert>C\cdot \lambda$.}
	\label{fig:4}	
\end{figure*}

\subsection*{Risk assessment}

As mentioned above, the crossing of particle in between super-structure provides a possible route for long-distance air transmission between the two occupants (see Fig.~\ref{fig:4}A), and thus leads to the enhanced infection for respiratory diseases. To quantify this process, we calculate the frequency for particles to cross the large-scale airflow, i.e., $f=\frac{1}{T}\frac{1}{N}\sum\limits_{i}A_{i}(C)$, where $N$ is the total number of particles, $T$ the total evalution time and $A_{i}(C)$ the counts of crossing over multiple heat sources of number $C$ for i-th particle. Figure \ref{fig:4}B shows the frequency of particles crossing as a function of $L/\lambda$ for various $C$. It is seen that when the distance $\lambda$ exceeds a critical value ($L/\lambda_c = 4$), there is a significant enhancement in the frequency compared to the cases in regime of individual effect. This finding suggests that the reduction of the distance between occupants results in a vast increase of particle crossing, which in turn induces a high infection risk. 
By taking advantage of the morphological transition of the flow structure, we can minimize the risk of disease transmission by adjusting the spacing between heat-carrying bodies. This can help to inform the development of effective strategies for controlling the spread of infectious diseases in indoor environments.

\section*  {Discussion}
Our direct numerical simulations of airflow patterns and aerosol transport in poorly-ventilated rooms reveal that arranging occupants at a reasonable distance is crucial to minimize the transmission of potentially infectious aerosols. Accommodating a minimal number of people by ultizing the morphological phase transition of airflow in a room is the most effective way to reduce cross-contamination. 

Our study identified two regimes of thermal plume behavior from heating sources at different distancing: the individual regime and collective regime. In the individual regime, where distancing between occupants is large, thermal plumes act as ``thermal armour" and exhibit significant individual effects, limiting the spread of virus particles. In contrast, in the collective regime, where the number of occupants is increased, thermal plumes condense and merge into the super-structure, enhancing the intensity of airflow and facilitating long-distance transport of aerosols. Our study provides decision-makers with guidelines for determining the maximum number of occupants based on the underlying physics of viral transmission in enclosed spaces.

Our findings is especially important when normal indoor activities are resumed and one relies on the natural way of precautions based on the flow physics of the indoor flow. It also provides guidances on coping with the potential outbreak of respiratory diseases other than SARS-CoV-2 in the future.
    
\section* {Materials and Methods}
In the current study, we conducted direct numerical simulations (DNS) of three-dimensional (3D) turbulent convection with spatially harmonic heating. We consider the coupled equations of motion for the velocity field $\textbf{u}$ and the temperature field $T$ for convective flows in a poorly-ventilated room with various heating sources regularly arranged below. 
Under the Boussinesq approximation. the governing equations are 
\begin{gather}
\nabla \cdot \boldsymbol{u} = 0, \label{eq:01} \\
\partial_t \boldsymbol{u}+(\boldsymbol{u}\cdot\nabla)\boldsymbol{u}=-\frac{1}{\rho}\nabla p+ \nu \nabla^2\boldsymbol{u}+ \alpha g T \boldsymbol{e}_z, \label{eq:02} \\
\partial_t T + (\boldsymbol{u}\cdot\nabla) T =\kappa_T \nabla^2 T, \label{eq:03}
\end{gather}
where $t$ is the time, $\boldsymbol{u}$ is the fluid velocity, $\rho$ is the fluid density, $T$ is the temperature, and $\boldsymbol{e}_z$ is the vertical unit vector. 

The governing equations are numerically solved by DNS using the multi-scalar second-order finite-difference method with a fractional third-order Runge–Kutta scheme, which has been validated many times in the literature\cite{Verzicco_jcp_1996,Ostilla_jcp_2015}. At all solid boundaries, no-slip boundary conditions are applied for the velocity.
For boundary conditions, we apply a distributed heating to the bottom plate:

\begin{equation}
T_\mathrm{bot}(x)=
\begin{cases}

T_{ocpt},&\mbox{for  $ (n-1/2) \lambda \le x \le (n-1/2)\lambda + l_{ocpt}$ }\\

T_{amb},&\mbox{for others}.

\end{cases}
\end{equation} 
where $n = 1, 2, \cdots, k$, $l_{ocpt} = H/8$ is the width of occupants. A constant temperature $T_\mathrm{top} = T_{amb}$ holds on the top plate, and an insulating condition is adopted on the sidewalls. About the initial conditions, the temperature field is constant set to ambient temperature. And the velocity, the systems are motionless initially, without any perturbation. For each run, the uniform grid spacing is adopted horizontally while stretched grid spacing is adopted vertically to guarantee more grid points near the top and bottom walls for resolving small scales inside boundary layer. We employ a grid resolution to adequately resolve both the Kolmogorov length scale for the velocity field $\boldsymbol{u}$ and the Batchelor length scale for the temperature field $\theta$. For the temporal resolution, the time step adopted is set to be small enough to guarantee the numerical stability and resolve the Kolmogorov time scale of turbulent fluctuations.

For aerosols, we apply the spherical point-particle model. Considering that the majority of exhaled aerosols are smaller than $5 \mu m$\cite{Wangsic_sci_2021,Sima_ast_2020}, even a large fraction is $<1 \mu m$ for most respiratory activities, including those produced during breathing, talking, and coughing, we investigate the effect of heat sources on virus particle transmission from the Lagrangian viewpoint by tracking fluid particles and calculating particle statistics. Our focus here is on transmission via aerosols, which are small enough (and noninertial) that they can be regarded as faithful tracers of the fluid flow \cite{Mathai_sciacv_2021,Microbubbles_prl_2016,Warhaft_arfm_2000}. Further details of numerical parameters are given in the Supplementary Materials.

\bibliographystyle{ScienceAdvances}

\section*{Acknowledgments}
\textbf{Funding:} This work was supported by the Natural Science Foundation of China under Grant Nos. 11988102, 92052201, 91852202, 11825204, 12032016, 12102246, 91952102 and 11972220, the Shanghai Science and Technology Program under Project No. 20ZR1419800, the Shanghai Pujiang Program under grant No. 21PJ1404400.  \textbf{Author contributions:} K. L. C. conceived and designed research. C. -B. Z. and J.-Z. W. conducted the numerical simulations. K. L. C., B.-F. W., T. C., Q. Z. supervised the project. C.-B.Zhao, K. L. Chong and J.-Z. Wu wrote the manuscript. All authors contributed to the analysis of the data. \textbf{Competing interests:} The authors declare that they have no competing interests. \textbf{Data and materials availability:} All data needed to evaluate the conclusions in the paper are present in the paper. Additional data related to this paper may be requested from the authors.

\end{document}